\documentclass[aps,prb,groupedaddress,twocolumn,amsmath]{revtex4}

\usepackage{amsmath}
\usepackage{color}
\newcommand{\lhf}{\ensuremath{\mathrm{LiY_{1-x}Ho_xF_4}}}

\begin{document}

\title{Quantum corrections to the Ising interactions in \lhf}
\author{A. Chin and P. R. Eastham}
\affiliation{T.C.M., Cavendish Laboratory, J J Thomson Avenue, Cambridge. CB3 0HE. U. K.}

\date{\today}

\begin{abstract}
We systematically derive an effective Hamiltonian for the dipolar
magnet \lhf, including quantum corrections which arise from the
transverse dipolar and hyperfine interactions. These corrections are
derived using a generalised Schrieffer-Wolff transformation to leading
order in the small parameters given by the ratio of the interaction
energies to the energy of the first excited electronic state of the
Holmium ions. The resulting low-energy Hamiltonian involves two-level
systems, corresponding to the low-lying electronic states of the
Holmiums, which are coupled to one another and to the Holmium
nuclei. It differs from that obtained by treating the electronic
states of the Holmium as a spin-1/2 with an anisotropic g-factor. It
includes effective on-site transverse fields, and both pairwise and
three-body interactions among the dipoles and nuclei. We explain the
origins of the terms, and give numerical values for their strengths.
\end{abstract}


\maketitle

\section{Introduction}

The rare-earth compound \lhf has often been described as a model
quantum magnet, and as such has been studied for over three
decades\cite{cooke75,reich90,bitko96,giraud01,S.Ghosh06212002,ghosh03}.
The basic model for this material is a diluted Ising model with dipole
interactions, and depending on the dilution the low-temperature phase
is expected to be either a ferromagnet or a
spin-glass\cite{stephen81}. Applying a strong transverse field to the
material introduces quantum fluctuations which can lead to domain wall
tunnelling in the ferromagnet\cite{Brooke01}, and to quantum melting
of the ferromagnet\cite{bitko96} and the
spin-glass\cite{wu91,schechter05,schechter2006,gingras2006,schecterpreprint}. Even
in the absence of an applied transverse field, however, quantum
tunnelling of magnetic dipoles can be observed in \lhf
\cite{giraud01,giraud03}.  Furthermore, samples with a Holmium
concentration of $4.5\%$ do not appear to behave as a spin-glass on
cooling, as expected for a classical Ising model. Instead they show
unexplained ``antiglass'' properties which have been attributed to
quantum mechanics.\cite{S.Ghosh06212002} This attribution is supported
by theoretical predictions of the static susceptibility, which agree
with experiments only once quantum corrections are
included\cite{ghosh03}.

Although quantum effects are implicated in several phenomena in \lhf
in the absence of an applied transverse field, there is no derivation
in the literature of an effective low-energy Hamiltonian which
contains them. Here we derive such a Hamiltonian, projecting out the
high-energy electronic states to obtain a theory describing the
low-lying electronic doublet and the nuclei. We shall find a
Hamiltonian which has a different form from that appropriate in a
strong applied field\cite{chakraborty04}, and whose electronic part
differs from the two-level model previously proposed for the
zero-field case\cite{ghosh03}.

\section{Model}

The magnetic degrees of freedom in \lhf are the $f$ electrons on the
$\mathrm{Ho^{3+}}$ ions. The strong spin-orbit coupling of the ions
leads to a well-defined $J=8$ for the ions, and an associated dipole
moment $\mu=g_{L} \mu_B$, with the Land\'e g-factor $g_{L}=5/4$. The
$2J+1=17$-fold degeneracy of the free ion is broken by the
crystal-field Hamiltonian, leaving a degenerate ground-state doublet.
All matrix elements of $J_{x}$ and $J_{y}$ are zero within this
ground-state subspace, but there are non-zero matrix elements for
$J_{z}$. This is the source of the strong Ising anisotropy in the
interactions and response to an applied field.

The interactions between Holmium ions are dipolar, giving an
interaction Hamiltonian \begin{equation} H_{\mathrm{int}}= \frac{1}{2}
  \sum_{i,j} \frac{\mu_0 \mu_B^2 g_L^2}{4\pi
    \mathbf{r}_{ij}^3}(\mathbf{J}_i.\mathbf{J}_j -
  3(\mathbf{\hat{r}}_{ij}.\mathbf{J}_i)(\mathbf{\hat{r}}_{ij}.\mathbf{J}_j)).
\label{hint}\end{equation} The crudest way to obtain a low-energy effective Hamiltonian
from (\ref{hint}) is to truncate the electronic state-space to the
ground-state doublet on each Holmium ion.  It is possible to choose a
basis for the doublet in which $J_z$ has no off-diagonal matrix
elements, while the diagonal matrix elements are $\alpha$ and
$-\alpha$. With this choice of basis, truncating leads to the dipolar
Ising model
\begin{equation} H_{\mathrm{int}}=\frac{1}{2} \sum_{i,j} \frac{\alpha^2 \mu_0 \mu_B^2 g_L^2}{4\pi
    \mathbf{r}_{ij}^3}(1-3(\mathbf{\hat{r}}_{ij}.\mathbf{\hat{k}})^2)\sigma^z_i\sigma^z_j,\label{isingdipoles}\end{equation}
where $\sigma^x,\sigma^y$ and $\sigma^z$ are the usual Pauli matrices. As
discussed in previous studies of this
system\cite{schechter05,chakraborty04,bitko96,giraud01} the contact
hyperfine interaction between the Holmium nuclei and the electronic
degrees of freedom also plays an important role, as it is typically of
similar strength to the dipolar interaction. Under the simple
two-state truncation scheme described above the hyperfine interaction
also takes a simple Ising form,

\begin{eqnarray}
H_{\mathrm{hyp}}&=&\sum_{i}A_{J}\mathbf{I}_{i}\cdot\mathbf{J}_{i}
\label{eq:hhypf} \\ && \longrightarrow\alpha\sum_{i}A_{J}
I_{i}^{z}\sigma_{i}^{z}, \label{eq:hhypftruncated}
\end{eqnarray}
as do other couplings to the Holmium moments.

The procedure of projecting into the low-energy subspace of the
single-ion Hamiltonian has been applied in the presence of an applied
transverse field.\cite{chakraborty04} The field changes the low-energy
states, leading to finite matrix elements for $J_{x}$ and $J_{y}$
within the low-energy doublet. This leads to corrections to the Ising
forms (\ref{isingdipoles},\ref{eq:hhypftruncated}) that are pair interactions
involving at least one of $\sigma^{x},\sigma^{y},I^x,I^y$, and to
effective field terms.

To obtain quantum terms in the absence of an applied transverse field
we must go beyond a simple projection onto the low-lying single-ion
states. We thus now consider the lowest three levels of the
crystal-field Hamiltonian, denoting the two states of the doublet as
$|\uparrow\rangle$, $|\downarrow\rangle$ and the first excited state,
with energy $\Delta$, as $|\Gamma\rangle$.  The parameters in such a
three-level model are $\Delta$ and the matrix elements of the angular
momentum operators, and can be obtained from a numerical
diagonalisation of the full crystal field Hamiltonian. With an
appropriate choice of basis this gives, \cite{chakraborty04}
\begin{eqnarray} \langle\uparrow|J_z|\uparrow\rangle &=& -
\langle\downarrow|J_z|\downarrow\rangle = 5.52 = \alpha 
\label{eq:m1}, \\
\langle\downarrow|J_x|\Gamma\rangle &=&
\langle\uparrow|J_x|\Gamma\rangle = 2.4  = \rho  \label{eq:m2}, \\
\langle\downarrow|J_y|\Gamma\rangle &=& -
\langle\uparrow|J_y|\Gamma\rangle = i \rho  \label{eq:m3}, \\
\Delta&=&10.8\mathrm{K}. \label{eq:gapparam}
\end{eqnarray} All other matrix elements of angular
momentum among the three states vanish. 

\section{Derivation of the effective Hamiltonian}

Quantum corrections to the Ising Hamiltonian
(\ref{isingdipoles},\ref{eq:hhypftruncated}) appear once one includes
the state at $|\Gamma\rangle$ because the interaction terms
(\ref{hint}) and (\ref{eq:hhypf}) couple $|\Gamma\rangle$ to the
electronic ground-state doublet. The typical nearest-neighbour dipolar
interaction energy scale is $\approx 300\mathrm{mK}$ and $A_{J}\approx
38\mathrm{mK}$\cite{giraud01,chakraborty04}. Since these interaction
scales are small compared with $\Delta$ the couplings to the
$|\Gamma\rangle$ states can be treated in perturbation theory, and an
effective Ising model obtained using the standard Schrieffer-Wolff
procedure\cite{schrieffer66}. The resulting model will include quantum
terms arising from virtual transitions between the ground-state
doublet and the $|\Gamma\rangle$ state.

To derive the effective Hamiltonian we begin by writing the three-level
model as $H=H_{0}+H_{T}$, where $H_{0}$ contains terms which do not
couple the doublet to $|\Gamma\rangle$, and $H_{T}$ contains those
which do. Measuring energy from the electronic ground-state doublet,
$H_0$ contains the crystal field term $V_{c}=\Delta\sum_{i}
|\Gamma_{i}\rangle\langle\Gamma_{i}|$ and the Ising parts of the
interactions, while $H_{T}$ contains the parts of the interactions
which involve $J_{x}$ and $J_{y}$. Including the hyperfine interaction
the Hamiltonian can be spilt as \begin{equation}
H_{0}=V_{c}+\frac{1}{2}\sum_{i\neq
j}J_{ij}^{zz}J_{i}^{z}J_{j}^{z}+\sum_{i}A_{J}I_{i}^{z}J_{i}^{z},
\end{equation} \begin{equation}
H_{T}=\frac{1}{2} \sideset{}{'}\sum_{\substack{i\neq j \\ \nu,\mu}} J_{ij}^{\nu\mu}J_{i}^{\nu}J_{j}^{\mu}+A_{J}\sum_{i}(I_{i}^{x}J_{i}^{x}+I_{i}^{y}J_{i}^{y}),
\end{equation} where $\nu,\mu = x,y,z$, but the prime on the sum indicates that we must exclude all terms that have $\nu=\mu=z$.

We then seek a unitary transformation $H\to e^{S}He^{-S}$ which
decouples the electronic doublet from the $|\Gamma\rangle$ state to
first order in $H_{T}$. Such a transformation obeys
\begin{equation}
[S,H_0]=-H_{T}. \label{SW}
\end{equation} An $S$ which approximately satisfies (\ref{SW}) can
be constructed from $H_T$ using projection operators,
\begin{eqnarray} S&=&\sideset{}{'}\sum_{\substack{i \neq
j \\ \nu,\mu}} \frac{J_{ij}^{\nu\mu}(\Gamma_{i}^{\nu}\Gamma_{j}^{\mu}J_{i}^{\nu}J_{j}^{\mu}P_{i}P_{j}-
P_{i}P_{j}J_{i}^{\nu}J_{j}^{\mu}\Gamma_{i}^{\nu}\Gamma_{j}^{\mu})}{2\epsilon_{\nu\mu}\Delta}\nonumber\\
&+&\frac{A_{J}}{\Delta}\sum_{i}\Gamma_{i}(I_{i}^{x}J_{i}^{x}+I_{i}^{y}J_{i}^{y})P_{i}\nonumber\\
&-&\frac{A_{J}}{\Delta}\sum_{i}P_{i}(I_{i}^{x}J_{i}^{x}+I_{i}^{y}J_{i}^{y})\Gamma_{i},\label{S}
\end{eqnarray} \begin{eqnarray}
P_{i}&=& |\uparrow_{i}\rangle\langle\uparrow_{i}| +|\downarrow_{i}\rangle\langle\downarrow_{i}|,\\
\Gamma_{i}^{\nu} &=&\left\{ \begin{array}{ll}
                       |\Gamma\rangle\langle\Gamma |  & \mbox{if $\nu=x,y$}  \\
                       P_{i}  & \mbox{if $\nu =z$}
                    \end{array}
            \right.,\\
            \epsilon_{\nu\mu} &=&\left\{ \begin{array}{ll}
                       2  & \mbox{if $\nu=\mu$}  \\
                       1 & \mbox{if $\nu \neq \mu.$}
                    \end{array}
            \right. .
\end{eqnarray} The form (\ref{S}) for $S$ actually obeys the relation $[S,V_{c}]=-H_{T}$, and
although we have successfully eliminated the linear term in $H_{T}$
the remainder of the commutator $[S,H_{0}-V_{c}]$ generates new
couplings between the doublet and $|\Gamma\rangle$. However these
tunnelling terms are of order $(\mathrm{interaction})^{2}/\Delta,$ and
as they can only couple between the ground states and
$|\Gamma\rangle$, they can only contribute to the effective low-energy
Hamiltonian in higher order perturbation theory. Therefore they do not
contribute to the effective low-energy Hamiltonian to leading order in
$(\mathrm{interaction})^{2}/\Delta$, and vanish when we project onto
the Ising basis at the end of the Schrieffer-Wolff procedure.

Having eliminated $H_{T}$, the effective two-state Hamiltonian is
given to lowest order in $(\mathrm{interactions})^{2}/\Delta$
by \cite{schrieffer66}
\begin{equation}
H_{\mathrm{eff}} = \prod_{i}P_{i}(H_{0} +\frac{1}{2}[S,H_{T}])\prod_{i}P_{i}.
\label{heff}
\end{equation} This form extends the Ising interaction
Hamiltonian by including second-order processes in which $H_{T}$
  causes virtual transitions from the electronic doublet to
  $|\Gamma\rangle$ and then back again.

\section{Effective Hamiltonian}
Substituting the form (\ref{S}) for $S$ into Eq.\ \ref{heff} and
discarding irrelevant energy shifts we find that the effective
Hamiltonian takes the form \begin{equation}
H_{\mathrm{eff}}=H_{\mathrm{0}}+H_{\mathrm{D}}+H_{\mathrm{DTB}}+H_{\mathrm{N}}+H_{\mathrm{ND}}. \label{eq:heff}
\end{equation} This Hamiltonian operates in the space with sixteen states
for each ion: two low-lying electronic states, each with the eight
nuclear states. The various contributions to $H_{\mathrm{eff}}$ are
labelled according to the interactions that give rise to them.

\subsection{Dipolar processes}

The first term in (\ref{eq:heff}) arises from the second-order dipolar
processes that involve only single pairs of spins, and is
\begin{equation}
H_{\mathrm{D}}=\sum_{i}(h_{i}^{x}\sigma_{i}^{x}+h_{i}^{j}\sigma_{i}^{y})+\sum_{i\neq
j}\Delta_{ij}^{\nu\,\mu}\sigma_{i}^{\nu}\sigma_{j}^{\mu}.
\label{dipolar}
\end{equation} We see that the pairwise dipolar interaction thus generates both transverse interaction terms and effective magnetic fields. In appendix\ \ref{ap1} we give expressions for the strengths of the parameters
$h_{i}$ and $\Delta^{\nu\mu}$ in terms of the parameters of the
underlying three-level model. Using these relations we estimate the
characteristic magnitude of these corrections by calculating the
contribution from a single nearest-neighbour. This yields
$h_{x}\approx 3\mathrm{mK}, h_{y}\approx 0$ and $
\Delta^{\nu\mu}\approx 0.06\mathrm{mK}$, which may be compared to the
dipolar Ising energy $J_{NN}^{zz}\approx300\mathrm{mK}$, and the Ising
hyperfine splitting $\alpha A_{J}\approx 210\mathrm{mK}$. Note that
terms such as (\ref{dipolar}) describe tunnelling between the
low-lying electronic states. It is important to stress that this is
not sufficient to generate tunnelling between the two electro-nuclear
ground states of $H_0$\cite{schechter05}.

In addition to the dipolar processes involving only a single pair of
Holmium ions there are processes involving three ions. In such a
process the interaction between one pair of spins, say $i,j$,
virtually excites the $j$th spin into $|\Gamma\rangle$, and the
dipolar interaction of this spin with the $k$th spin brings the $j$th
spin back into the Ising doublet. This generates two- and three- body
interaction terms in the effective theory,
\begin{eqnarray}
H_{\mathrm{DTB}}&=&-\frac{\alpha^{2}\rho^{2}}{\Delta}\sum_{i\neq j\neq
k}(J_{ik}^{xz}J_{ij}^{xz}+J_{ik}^{yz}J_{ij}^{yz})\sigma_{k}^{z}\sigma_{j}^{z}\\
&-&\frac{\alpha^{2}\rho^{2}}{\Delta}\sum_{i\neq j\neq
k}(J_{ik}^{xz}J_{ij}^{xz}-J_{ik}^{yz}J_{ij}^{yz})\sigma_{k}^{z}\sigma_{j}^{z}\sigma_{i}^{x}
\\ &-&\frac{\alpha^{2}\rho^{2}}{\Delta}\sum_{i\neq j \neq
k}(J_{ik}^{xz}J_{ij}^{yz}+J_{ik}^{yz}J_{ij}^{xz})\sigma_{k}^{z}\sigma_{j}^{z}\sigma_{i}^{y}.\end{eqnarray}
For an isolated group of three spins in which $\{ i,j \}$ and $\{ i,k\}$
are nearest neighbours we obtain 3 mK for the magnitude of the
interaction terms $\sigma_k^z \sigma_j^z,
\sigma_{k}^{z}\sigma_{j}^{z}\sigma_{i}^{x}$ and $
\sigma_{k}^{z}\sigma_{j}^{z}\sigma_{i}^{y}$.

\subsection{Hyperfine processes}

The contributions to the effective Hamiltonian from the hyperfine
interactions are simpler to deal with, as they are confined to each
site. They are
\begin{eqnarray}
H_{\mathrm{N}}&=&\frac{\rho^{2}A_{J}^{2}}{\Delta}\sum_{i}\left((I_{i}^{2}-(I_{i}^{z})^{2})+((I_{i}^{+})^{2}+(I_{i}^{-})^{2})\sigma_{i}^{x}\right)\nonumber\\
&-&\frac{\rho^{2}A_{J}^{2}}{\Delta}\sum_{i}\left(i((I_{i}^{+})^{2}-(I_{i}^{-})^{2})\sigma_{i}^{y}-2I_{i}^{z}\sigma_{i}^{z}\right).
\end{eqnarray}

We see that in second order perturbation theory the hyperfine
interaction leads to spin flipping terms which require the z component
of the nuclear spin to change by two units, gives a slight correction
to the longitudinal hyperfine interaction, and introduces a weak
nuclear anisotropy. The factors outside the sum give $0.86
\mathrm{mK}$ for the interaction energy.

\subsection{Mixed hyperfine-dipolar processes}
Finally, the term in the effective Hamiltonian labelled
$H_{\mathrm{ND}}$ arises due to second-order processes whereby the
electrons are virtually excited into the state $|\Gamma\rangle$ by the
dipolar interaction and then de-excited by the hyperfine interaction
or vice versa:
\begin{eqnarray} H_{\mathrm{ND}}&=&-\frac{\alpha\rho^2 A_{J}}{\Delta}\sum_{i\neq
j}J_{ij}^{zx}(I_{i}^{x}(1+\sigma_{i}^{x})+I_{i}^{y}\sigma_{i}^{y})\sigma_{j}^{z}\nonumber\\
&-&\frac{\alpha\rho^2 A_{J}}{\Delta}\sum_{i\neq
j}J_{ij}^{zy}(I_{i}^{y}(1-\sigma_{i}^{x})+I_{i}^{x}\sigma_{i}^{y})\sigma_{j}^{z}.
\end{eqnarray} Note that the nuclei mediate a coupling between
electronic states of the form $\sigma^{x,y}\sigma^{z}$ which does not
occur in $H_{\mathrm{D}}$.

\section{Discussion}

In recent
papers\cite{schechter2006,gingras2006,schechter05,schecterpreprint} it
was shown that the non-Ising parts of the dipolar coupling provide a
route by which an applied transverse field can destroy spin-glass
order in \lhf. Although our model contains spontaneous transverse
fields we do not expect them to destroy the spin-glass phase in this
way, because the model retains time-reversal symmetry. An interesting
feature of the effective Hamiltonian is that the spontaneous
transverse fields and the two- and three- body interactions are all
correlated. Theoretical work on spin models with several forms of
random couplings, such as the random transverse-field Ising
model\cite{liu06}, generally considers different interaction terms as
independent.

The approach given here could be straightforwardly extended to derive
an effective Hamiltonian for \lhf in an applied transverse field, so
long as the field strength is sufficiently small for perturbation
theory to apply. This generates a variety of new processes at second
order. The straightforward process, in which the applied field excites
to the $|\Gamma\rangle$ state and back again, leads to an electronic
tunnelling term $\propto B^2/\Delta$. But note that there will also be
mixed processes, for example where the applied field causes the
transition to $|\Gamma\rangle$ and the dipole interaction causes the
transition back into the electronic doublet, which will generate
electronic tunnelling terms $\propto B/\Delta$.

We stress that although our Hamiltonian contains terms which generate
transitions between the states of the electronic Ising doublet, and
further such terms will be introduced by an applied field, these terms
alone do not couple the doubly-degenerate electro-nuclear ground
states of the Ising single-ion Hamiltonian. To generate quantum
fluctuations between the electro-nuclear ground states requires
processes which flip the nuclear spin, and hence the splittings of the
electro-nuclear ground states will be much smaller than the tunnelling
terms connecting the bare electronic states.

Although we do not expect the quantum corrections to formally destroy
the ordered phase unless there is an applied field, they can still
affect the thermodynamics, changing the susceptibilities and moving
the phase boundary. However, we only expect this to occur at very low
temperatures. In this context we note that the experimental
susceptibility of \lhf in the 10 -- 100 mK temperature regime of the
antiglass experiment has been reproduced by a quantum theory of a
two-level model\cite{ghosh03}. That model is obtained by neglecting
the nuclei and treating the Holmium ions as spin-$1/2$ ions with an
anisotropic g-factor, \emph{i.e.}  writing $J^\mu_i=g^\mu
\sigma^\mu_i$, with $\mu={x,y,z}$. We note that the resulting
Hamiltonian is very different from that which would be obtained by
neglecting the nuclei in our model: there are no spontaneous field
terms and the non-Ising interactions decay as $1/r^3$ (here $1/r^6$),
while the quoted values for $g$ give the energy scale for the largest
non-Ising coupling of $\approx 30 \mathrm{mK}$, whereas here we have
$3 \mathrm{mK}$.

While we do not expect the small corrections derived here to affect
thermodynamics except at very low temperatures, they may be relevant
to understanding dynamics at much higher temperatures. Quantum
tunnelling of the magnetisation has been observed in the magnetisation
relaxation and susceptibility of the dilute compound
$\mathrm{LiY}_{0.998}\mathrm{Ho}_{0.002}\mathrm{F}_{4}$, due to both
single-ion\cite{giraud01} and two-ion processes\cite{giraud03}, and it
would be interesting to compare the details of these results with the
processes given here. An understanding of the single or few-ion
electro-nuclear dynamics, based on the Hamiltonian (\ref{eq:heff}),
may also help to explain the antiglass
experiments\cite{S.Ghosh06212002}. In the high-temperature regime
these experiments show a characteristic relaxation time for the
magnetisation which is activated, with a barrier similar to the width
of the hyperfine multiplet and the nearest-neighbour interaction
strength. Extrapolating this behaviour indicates that this particular
activated dynamics freezes out on the experimental frequency scale as
the temperature is lowered into the antiglass regime. The quantum
terms derived here, although small, may perhaps then provide a route
to the observed dynamics. The potential significance of quantum
electro-nuclear dynamics for the antiglass experiment has been noted
by Atsarkin\cite{atsarkin88}, who proposes a specific relaxation
mechanism due to interactions of the form (\ref{dipolar}).

\section{Conclusions}
In this paper we have motivated the need to go beyond the simple
dipolar electro-nuclear Ising Hamiltonian commonly used to describe
\lhf, and have derived an effective low energy Hamiltonian which
includes the quantum corrections caused by the transverse elements of
the dipolar and hyperfine interactions. We have given estimates for
the typical magnitudes of these correction terms and have shown that
they are typically about one percent as strong as the energy scale
associated with the Ising interactions. As a result, we do not expect
any qualitative changes to the Ising phase diagram, but as we have
highlighted in this paper, these quantum corrections can describe a
large variety of single and many-body processes which might play
significant roles in the observed dynamics. Thus, for low
temperatures, our effective Hamiltonian should serve as a good
starting point for a microscopic investigation of the dynamical
physics of \lhf, and as it can be used across the whole dilution
series, should contain the rich low energy physics which characterises
the spin glass, free ion, and presumably, the ``antiglass'' phases of
this material.

\acknowledgments We are grateful to Misha Turlakov, Peter Littlewood,
and to the participants of the summer school, for helpful and interesting
discussions of this problem.

\appendix
\section{Relation of microscopic parameters to the effective interaction and on-site terms in $H_{D}$}
\label{ap1}
The components of the effective magnetic field $\mathbf{h}_{i}$, defined in Eq. \ref{dipolar}, are related to the original microscopic Hamiltonian by,

\begin{eqnarray}
h_{i}^{x}&=&\frac{\rho^{2}\alpha^{2}}{\Delta}\sum_{j}((J_{ij}^{zy})^{2}-(J_{ij}^{zx})^{2})\nonumber\\
&+&\frac{\rho^{4}}{2\Delta}\sum_{j}((J_{ij}^{yy})^{2}-(J_{ij}^{xx})^{2}),\\
h_{i}^{y}&=&-2\frac{\rho^{2}\alpha^{2}}{\Delta}\sum_{j}J_{ij}^{zy}J_{ij}^{zx}\nonumber\\
&-&\frac{\rho^{4}}{2\Delta}\sum_{j}J_{ij}^{xy}(J_{ij}^{xx}+J_{ij}^{yy})\\
h_{i}^{z}&=&0.
\end{eqnarray}
The components of the effective magnetic field vanish in the undiluted crystal, as expected from the crystal symmetry.

The transverse dipolar interactions between spins are described by the following couplings,
\begin{eqnarray}
\Delta_{ij}^{xx}&=&\frac{\rho^{4}\left(2(J_{ij}^{xy})^{2}-(J_{ij}^{xx})^{2}-(J_{ij}^{yy})^{2}\right)}{4\Delta},\\
\Delta_{ij}^{yy}&=&\frac{-\rho^{4}\left(J_{ij}^{xx}J_{ij}^{yy}+J_{ij}^{xy}J_{ij}^{xy}\right)}{2\Delta},\\
\Delta_{ij}^{zz}&=&\frac{\rho^{4}\left(J_{ij}^{xx}J_{ij}^{yy}-J_{ij}^{xy}J_{ij}^{xy}\right)}{2\Delta},\\
\Delta_{ij}^{xy}&=&\frac{\rho^{4}\left(J_{ij}^{yy}J_{ij}^{xy}-J_{ij}^{xx}J_{ij}^{xy}\right)}{2\Delta},\\
\Delta_{ij}^{yx}&=&\Delta_{ij}^{xy}. \end{eqnarray} These are the only
couplings generated; there are no terms such as $\sigma^{x}\sigma^{z}$, since these would break time-reversal symmetry.


\end{document}